\documentclass[reprint,amsmath,amssymb,aps,prl]{revtex4-2}
\usepackage[utf8]{inputenc}
\usepackage{natbib}
\usepackage[english]{babel}
\usepackage{graphicx}
\usepackage{array}
\usepackage{xfrac}
\usepackage{hyperref}
\usepackage{xcolor}

\usepackage{mathtools}

\newcommand{\tpeak}{\ensuremath{t^\mathrm{peak}}}
\newcommand{\tpeaktilde}{\ensuremath{\tilde{t}^\mathrm{peak}}}
\newcommand{\xpeak}{\ensuremath{x^\mathrm{peak}}}

\begin{document}
\title{Dynamic perturbation spreading in networks}

\author{Malte Schröder}
\affiliation{Chair for Network Dynamics, Institute of Theoretical Physics and Center for Advancing Electronics Dresden (cfaed), TU Dresden, 01062 Dresden, Germany}
\author{Xiaozhu Zhang}
\affiliation{Chair for Network Dynamics, Institute of Theoretical Physics and Center for Advancing Electronics Dresden (cfaed), TU Dresden, 01062 Dresden, Germany}
\author{Justine Wolter}
\affiliation{Chair for Network Dynamics, Institute of Theoretical Physics and Center for Advancing Electronics Dresden (cfaed), TU Dresden, 01062 Dresden, Germany}
\author{Marc Timme}
\affiliation{Chair for Network Dynamics, Institute of Theoretical Physics and Center for Advancing Electronics Dresden (cfaed), TU Dresden, 01062 Dresden, Germany}

\hyphenation{op-tical net-works semi-conduc-tor}

\begin{abstract}
Understanding how local perturbations induce the transient dynamics of a network of coupled units is essential to control and operate such systems.
Often a perturbation initiated in one unit spreads to other units whose dynamical state they transiently alter. The maximum state changes at those units  and the timings of these changes constitute key characteristics of such transient response dynamics. 
However, even for linear dynamical systems it is not possible to analytically determine time and amplitude of the maximal response of a unit to a perturbation. 
Here, we propose to extract approximate peak times and amplitudes from effective expectation values used to characterize the typical time and magnitude of the response of a unit by interpreting the system's response as a probability distribution over time. We derive analytic estimators for the peak response based on these expectation value measures in linearized systems operating close to a stable fixed point. These estimators can be expressed in terms of the inverse of the system's Jacobian.
We obtain identical results with different approximations for the response dynamics, indicating that these estimators become exact in the limit of weak coupling. 
Furthermore, the results suggest that perturbations spread ballistically in networks with diffusive coupling.

\end{abstract}

\maketitle

\section{Introduction}\label{sec:introduction}

Transient collective dynamics plays an important role in a wide range of systems from social and biological systems where ideas or diseases spread \cite{gautreau07_arrivaltimesDisease, gautreau08_diseaseSpread, brockmann13_spreading, Iannelli17_effectiveDistances, kirst2016_informationRouting, chen18_arrivalTimesLinearSpreading} to the stability of large scale infrastructure and supply networks such as power grids \cite{witthaut16_critical, manik17_network_susceptibilities, kettmann16_acgrids, Menck14_powergrids}. 
These systems generically operate near a fixed point and are naturally subject to perturbations, for example an outbreak of an infection or fluctuations of the power consumption and production \cite{schafer18_powerFluctuations}. In their simplest setting, such perturbations initially affect only a single unit and spread through the network, transiently affecting other units at different times and with different strengths \cite{hens2018_predicting, timme2019_propagation_patterns}.

Despite the importance of these spreading and propagation processes, no general answer exists for when or how strongly a unit is affected by an initial perturbation. 
Traditional measures to characterize these transient responses are the time and magnitude of the maximal (peak) response. However, even in linearized systems, computing such measures typically involves the solution of transcendental equations making exact analytical predictions impossible. 

A recently introduced idea \cite{wolter18_expectationSpreading} is interpreting the deterministic transient responses as probability densities in time. The resulting ``effective expectation values'' constitute characteristic response measures (different from traditional ones) that are computable analytically in linearized systems in terms of the inverse effective coupling matrix.

In this article we derive analytic estimators for the peak response time and amplitude based on these expectation value response measures in linearized systems affected by perturbations around a stable fixed point. We approximate the response dynamics with multiple different functions that  qualitatively reproduce it. For each approximation function, we analytically derive both, the response strength and timing as calculated from the expectation values and the amplitude and timing of maximal response. Comparing these results we find analytic estimators for the actual peak response time and magnitude in terms of the inverse Jacobian of the linearized system. We illustrate that such estimators become exact in the limit of weak coupling, independent of the topology of the coupling network.

\begin{figure*}
\centering
\includegraphics[width=0.9\textwidth]{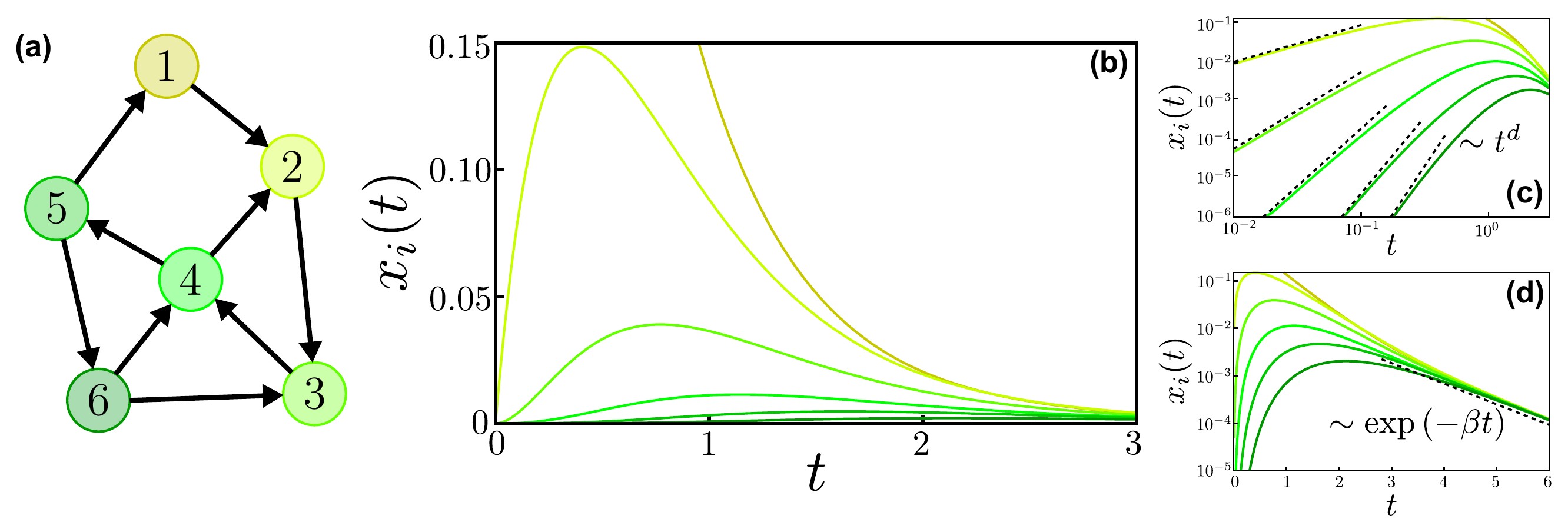}
\caption{\textbf{Typical response to a localized perturbation.} (a) A small network dynamical system with $N=6$ units and $E=9$ directed interactions. (b) Responses $x_i(t)$ of all units $i$ after an initial perturbation $x_1(0) = 1$ at unit $k = 1$ (dark yellow) with homogeneous coupling strength $\alpha = 1$ and internal dynamics $\beta = 1$. The perturbation spreads through the network and transiently affects all other units before the system returns to the stable operating state $x_i^* = 0$ for $t \rightarrow \infty$. 
(c,d) For small times $t \rightarrow 0$ the activity $x_i(t)$ of unit $i$ grows polynomially as $t^d$ with an exponent $d = i - 1$, the graph-theoretical distance from the initially perturbed unit. For large times $t \rightarrow \infty$ the activity of all units decays exponentially with exponent $-\beta$.}
\label{fig:example}
\end{figure*}

\section{Perturbations in\\ network dynamical systems}

Consider a general network dynamical system 
\begin{equation}
\frac{\mathrm{d}\mathbf{y}}{\mathrm{d}t}=\mathbf{F}(\mathbf{y})
\label{eqn:nonlinearsystem_general}
\end{equation}
consisting of $N$ coupled units $i$ with internal state $y_i(t) = \left[\mathbf{y}(t)\right]_i$ operating close to a stable fixed point $\mathbf{y}^* \in \mathbb{R}^N$. Small perturbations to this state and their impact across the network (Fig.~\ref{fig:example}) are described by the linearized dynamics 
$\mathrm{d}\mathbf{x}/\mathrm{d}t = M\mathbf{x}$ where $\mathbf{x}(t) = \mathbf{y}(t)-\mathbf{y}^*$ and $M = \mathrm{d}\mathbf{F}/\mathrm{d}\mathrm{y}\mid_{\mathbf{y}=\mathbf{y}^*}$ is the Jacobian. The diagonal elements of $M$ describe the internal dynamics of the individual units while the off-diagonal elements describe the coupling between the units.\\

In general, the impact of a perturbation on a unit and how a perturbation spreads through the network can be measured in different ways: for models of epidemic spreading, describing an outbreak across different populations coupled by a transportation network, arrival times are often defined by measuring the first time $t_\mathrm{arrival} = \mathrm{min}\left\{t : x_i(t) \ge \epsilon \right\}$ when the number of infected individuals exceeds a given threshold $\epsilon$. For stochastic epidemic spreading the connection to random walk processes allows predictions of the arrival times of the perturbation \cite{braunstein03_optimalpaths, gautreau07_arrivaltimesDisease, roosta82_reliableRouting, brockmann13_spreading, Iannelli17_effectiveDistances}.
For other spreading processes the total impact of the perturbation at a given unit or the maximal deviation from the operating point is of interest \cite{hens2018_predicting, kittel17_timing, poolla2017optimal, tyloo18_robustness_kirchhoff, hellmann16_survivability, timme2019_propagation_patterns}. Interestingly, even for the simple deterministic linear system $\mathrm{d}\mathbf{x}/\mathrm{d}t = M\mathbf{x}$ described above, these measures cannot be easily evaluated analytically. The underlying reason is that calculating peak positions or threshold crossing times typically involves solving transcendental equations of the form $a_1 \exp\left(\lambda_1 t\right) + a_2 \exp\left(\lambda_2 t\right) + \ldots = c$ for the time $t$, where $\lambda_j$ are the eigenvalues of $M$.

Recently, a complementary approach was introduced to characterize the impact of a perturbation in such linearized systems in terms of expectation values of effective probability distributions \cite{wolter18_expectationSpreading}. This approach works as follows: We first normalize each response trajectory and interpret the result as a probability density over time. We then quantify the arrival time and impact of a perturbation by expectation values and higher order moments with respect to this effective probability distribution.

\begin{figure*}
\centering
\includegraphics[width=0.9\textwidth]{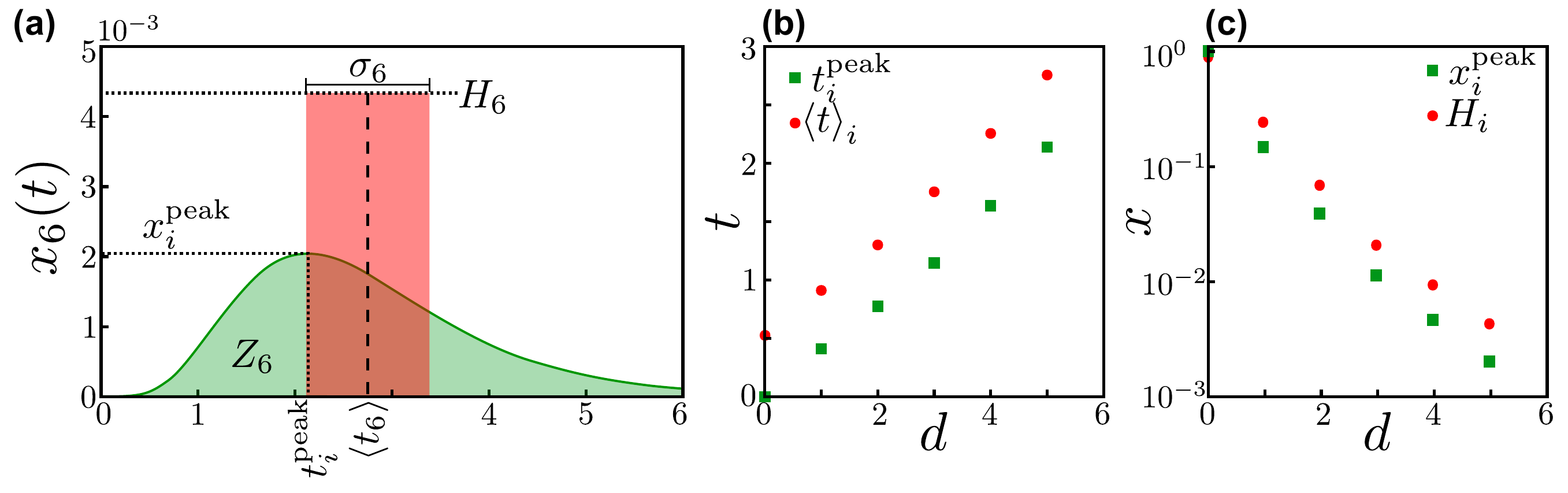}
\caption{\textbf{Expectation values quantify the relative impact of the perturbation.} (a) Interpretation of the response trajectories as probability densities over time yields the characteristic response measures introduced in \cite{wolter18_expectationSpreading} based on the effective expectation values defined in Eq.~(\ref{eq:total_response} - \ref{eq:height}). The example shows the response $x_6(t)$ of unit $6$ (dark green, compare Fig.~\ref{fig:example}) and the corresponding characteristic response measures (red). (b,c) These measures accurately characterize the \emph{relative} impact of the perturbation and show the same scaling as the actual peak response times and magnitudes, illustrated here for the small example shown in Fig.~\ref{fig:example} where $d = i - 1$ is the graph-theoretical distance from the perturbed unit. The response times seem to be biased additively, the response strengths multiplicatively. This observed systematic difference suggests that an adjustment is possible to obtain estimators for the \emph{absolute} peak response values based on the characteristic (expectation value) response measures.}
\label{fig:expectation}
\end{figure*}

In the following we assume that the internal dynamics of each unit is described by exponential decay with a rate $\beta_i > 0$ ($\mathbf{x} = \mathbf{0}$ is a stable fixed point) and the coupling between the units is diffusive with coupling strength $\alpha_{ij} \ge 0$, such that $M_{ij} = \alpha_{ij}$ (for $i \neq j$), $\alpha_{ij} = 0$ if unit $i$ is not directly affected by unit $j$ and $M_{ii} = -\beta_i - \sum_j \alpha_{ij}$. In this case the response of each unit $x_i(t)$ to an initial perturbation $x_i(0) = \left(\mathbf{x}_0\right)_i = \delta_{ik}$ at a single unit $k$ is guaranteed to be positive, $x_i(t) > 0$ for all times $t > 0$ (see Fig.~\ref{fig:example}). 
Appropriate normalization of the response trajectories $x_i(t) = \left[\exp\left(Mt\right)  \mathbf{x}_0\right]_i$ by the total response
\begin{equation}
Z_i = \int_0^\infty x_i(t) \mathrm{d}t = -\left( M^{-1} \mathbf{x_0} \right)_{i} \label{eq:total_response}
\end{equation}
then allows the interpretation of the trajectories as if they were probability densities over time, $\rho_i(t) = x_i(t) / Z_i$. From this perspective, expectation values of time with respect to the probability distribution characterize the impact of the perturbation at different units (see Fig.~\ref{fig:expectation}) with simple analytic expressions \cite{wolter18_expectationSpreading}. For example, the expectation value
\begin{equation}
\left<t\right>_i = \int_0^\infty t \rho_i(t) \mathrm{d}t = -\frac{\left( M^{-2} \mathbf{x_0} \right)_{i}}{\left( M^{-1} \mathbf{x_0} \right)_{i}}	\label{eq:response_time}
\end{equation}
describes the characteristic response time (not the peak time) when the perturbation impacts unit $i$. 
Similarly, the typical duration of the perturbation is measured in terms of the standard deviation $\sigma_i$ and its magnitude by the quotient $H_i$ of total response $Z_i$ and the standard deviation
\begin{eqnarray}
	\sigma_i &=& \sqrt{\int_0^{\infty}  (t-\left<t\right>_i)^2 \rho_i(t) dt} \label{eq:duration}\\
		&=& \sqrt{ \frac{2(M^{-3} \mathbf{x}_0)_i}{(M^{-1} \mathbf{x}_0)_i} - \left(\frac{(M^{-2} \mathbf{x}_0)_i}{(M^{-1} \mathbf{x}_0)_i}\right)^2 }	\nonumber\\
	H_i &=& \frac{Z_i}{\sigma_i} = \frac{((M^{-1} \mathbf{x}_0)_i)^2}{\sqrt{2(M^{-3} \mathbf{x}_0)_i (M^{-1} \mathbf{x}_0)_i- ((M^{-2} \mathbf{x}_0)_i)^2}}	\,.\label{eq:height}
\end{eqnarray}
These quantities are illustrated in Fig.~\ref{fig:expectation} together with the numerically determined peak response values $\tpeak_i$ and $\xpeak_i = x_i(\tpeak_i)$ for the example system from Fig.~\ref{fig:example}. As also demonstrated previously \cite{wolter18_expectationSpreading}, the characteristic response times and response magnitudes [Eq.~(\ref{eq:response_time}) and (\ref{eq:height})] appear to show the same scaling as the actual peak time $\tpeak_{i}$ and the maximal response $x_i(\tpeak_{i})$, that means they accurately describe the \emph{relative} impact of the perturbations at different units. However, if interpreted as estimators for the peak response values they are clearly biased and do not provide a good quantitative description of the \emph{absolute} impact. 

In general, for unimodal distributions as we observe for the typical response trajectories, some conditions on the relationship between mean (expectation value) and mode (position of the maximum) are known. For example, if the distribution has positive skewness (as the response trajectories), we typically have $\left<t\right>_i \ge \tpeak_i$. Unimodal distributions also satisfy the condition $\left| \tpeak_i - \left<t\right>_i \right| \le \sqrt{3} \sigma_i$ \cite{johnson51_moment}. Unfortunately, no exact connection between mean and mode for general distributions, and thereby for general $x_i(t)$, exists.

For the specific class of (initially algebraically increasing and then exponentially decaying) responses $x_i(t)$ in linearized systems, we here establish two connections between the characteristic response values and the actual peak values. Specifically, the examples in \cite{wolter18_expectationSpreading} already suggest that the relation between the actual peak measures and the characteristic response measures is systematic and largely independent of the structure of the interaction network, as  also illustrated in Fig.~\ref{fig:expectation}(b,c). As illustrated in Fig.~\ref{fig:expectation}(b), we observe an approximately constant shift between $\left<t\right>_i$ and $\tpeak_i$. This suggests an additive adjustment $c_T$ to estimate the actual peak time
\begin{equation}
	\mathrm{Est}\left[\tpeak\right]_i = \left<t\right>_i - c_T \,.
\end{equation}
Similarly, as illustrated in Fig.~\ref{fig:expectation}(c) we also observe a constant multiplicative factor between $H_i$ and $\xpeak_i$ (note the logarithmic axis), suggesting a multiplicative adjustment $c_H$ 	such that 
\begin{equation}	
	\mathrm{Est}\left[\xpeak\right]_i = c_H \, H_i \,.
\end{equation}

In the following we analytically derive these adjustments from approximate response functions and show that they results in the same form of adjustment. We use these calculations to determine the constants $c_T$ and $c_H$ and define the estimators for the actual peak response based on approximating model trajectories.
For different classes of model functions that recover the qualitative shape and the asymptotic behavior of the response dynamics we calculate exact characteristic and peak response values as an explicit function of the interaction network described by $M$. 
We use these expressions to convert the characteristic response measures resulting from effective expectation values into estimators for the peak values for these approximating functions and thereby for the real response dynamics.

\section{Estimators for the peak response}

\begin{figure*}[ht]
\centering
\includegraphics[width=0.9\textwidth]{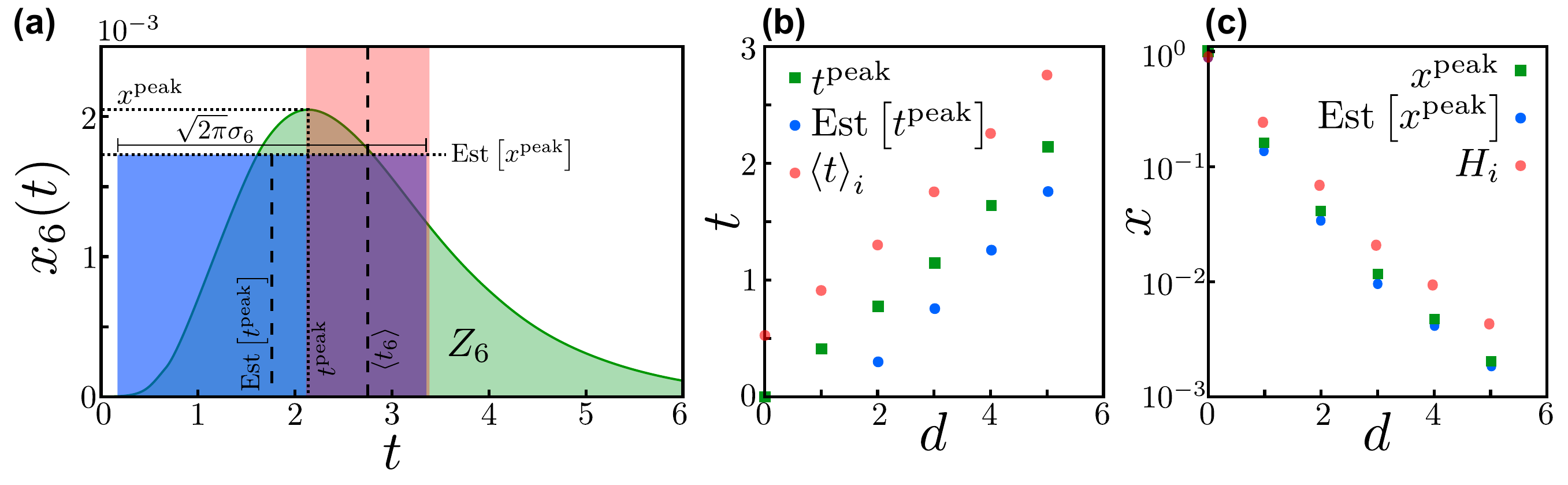}
\caption{\textbf{Adjusted expectation values.} (a) Illustration of the adjusted expectation value quantifiers (compare Fig~\ref{fig:expectation}). (b,c) With the adjustments derived in the main text the expectation values [Eq.~(\ref{eq:total_response} - \ref{eq:height})] are converted into more accurate estimators [Eq.~(\ref{eq:est_t}) and (\ref{eq:est_h})] for the actual time and especially the peak magnitude of the response.}
\label{fig:estimators}
\end{figure*}

The asymptotic behavior of the units' responses is given by polynomial growth for $t \rightarrow 0$ and by exponential decay for $t \rightarrow \infty$ [compare Fig.~\ref{fig:example}(c,d)]. 
Motivated by these known asymptotic scaling regimes, we illustrate the approach to calculate the constants $c_T$ and $c_H$ using a simple family of approximating functions 
\begin{equation}
	\tilde{x}(t) = \tilde{A} \, t^{\tilde{d}} \, \exp(-\tilde{\beta} t) \,\label{eq:approximation_family}
\end{equation}
that roughly capture the qualitative behavior of the response dynamics. Since the constant factor $\tilde{A}$ does not change the peak position or the factors $c_T$ and $c_H$, we set $\tilde{A} = 1$ in the following. We now first determine the remaining parameters $\tilde{d}$ and $\tilde{\beta}$ of this approximate response function.

To understand the asymptotic behavior of the units for small times, $t \rightarrow 0$, we consider the formal solution
\begin{eqnarray}
\mathbf{x}(t) &=& \exp(Mt) \mathbf{x}_0 \\
	&=& \mathbf{x}_0 + t M \mathbf{x}_0 + 1/2 \; t^2 M^2 \mathbf{x}_0 + \ldots \,.	\nonumber
\end{eqnarray}
For a perturbation at a single unit $k$ with $x_k(0) = 1$ this expression reduces to the matrix elements $x_i(t) = \delta_{ik} + t M_{ik} + 1/2 \; t^2 (M^2)_{ik} + \ldots\,$. The matrix $M$ is directly related to the adjacency matrix of the interaction network (with additional entries along the diagonal). Thus, for networks with homogenous coupling strengths $\alpha_{ij} = \alpha$, the element $(M^n)_{ik}$ is proportional to the number of paths from $k$ to $i$ of length $n$ \cite{xiaozhu18_thesis}. 
If we denote the (shortest path) distance from the initially perturbed unit $k$ to unit $i$ as $d \in \mathbb{N}$, all elements $(M^n)_{ik} = 0$ for $n<d$ since there are no paths of length $n < d$. This means that the first non-zero term in the response of unit $i$ is given by $x_i(t) = 1/d! (M^d)_{ik} t^d + \mathcal{O}\left(t^{d+1}\right)$ as $t \rightarrow 0$. The same argument holds for networks with heterogenous coupling strengths $\alpha_{ij}$. The entry $(M^n)_{ik}$ is then given by the sum over all weighted paths of length $n$.

For large times, $t \rightarrow \infty$, we consider the eigenvalues $-\lambda_i$ of $M$. We explicitly write them as $-\lambda_i$ to signify that all eigenvalues have negative real part $\mathrm{Re}\left[-\lambda_i \right] < 0$ since $M$ describes the dynamics around an asymptotically stable fixed point. We label the eigenvalues such that $\mathrm{Re}\left[-\lambda_N \right] \le \dots \le \mathrm{Re}\left[-\lambda_2 \right] \le \mathrm{Re}\left[-\lambda_1 \right]$. The response in terms of these eigenvalues is then given by
\begin{eqnarray}
x_i(t) = \left[\mathbf{x}(t)\right]_i &=& \left[\exp(Mt) \mathbf{x}_0\right]_i \\ \label{eq:eigenvalue_expansions}
 &=& c_1 \exp(-\lambda_1 t) + c_2 \exp(-\lambda_2 t) + \dots \nonumber \\
 &=& \exp(-\lambda_1 t) \left[ c_1 + c_2 \exp\left( \left(\lambda_1-\lambda_2\right) t\right) + \dots \right] \,, \nonumber
\end{eqnarray}
with constants $c_i$ depending on the initial conditions $\mathbf{x}_0$. For undirected networks (symmetric $M$) $c_j = \left(\mathbf{v}_j^T\mathbf{x}_0\right) \left[\mathbf{v}_j\right]_i$ where $\mathbf{v}_1, \mathbf{v}_2, \dots$ denote the orthogonal eigenvectors of $M$ corresponding to the (real) eigenvalues $-\lambda_1, -\lambda_2, \dots$. For large $t \rightarrow \infty$ the first term dominates. Thus, the response at all units is given by $x_i(t) = c_1 \exp(-\lambda_1 t) + \mathcal{O}\left[ \exp(-\lambda_2 t) \right]$ with magnitude $\mathcal{O}\left[\exp( \mathrm{Re}\left[-\lambda_1\right] t) \right]$. For notational convenience, we do the following calculations for undirected networks with real eigenvalues and drop the real part notation, writing only $-\lambda_1$.\\
 
Matching the above considerations for large and small $t$, defining the parameters of $\tilde{x}(t)$ as $\tilde{d} = d$ and $\tilde{\beta} = \lambda_1$, Eq.~(\ref{eq:approximation_family}) becomes
\begin{equation}
\tilde{x}(t) = t^d \exp( -\lambda_1 t) \,.
\end{equation}
For this approximation for the response of a unit at (shortest path) distance $d$ to the initial perturbation we now calculate both the typical response measures [Eq.~(\ref{eq:response_time}) and (\ref{eq:height})] as well as the true peak response values analytically. The normalization factor $\tilde{Z}$ [Eq.~(\ref{eq:total_response})] is
\begin{equation}
\tilde{Z} = \int_0^\infty \tilde{x}(t) \mathrm{d}t = \frac{d!}{\lambda_1^{d+1}}	\,,\label{eq:partial_integration_math}
\end{equation}
and we define $\tilde{\rho}(t) = \tilde{x}(t) / \tilde{Z}$. Here and in the following we drop the indices of the response function denoting the dependence on the unit $i$ (and the initially perturbed unit $k$). These dependencies become explicit by noting that the graph-theoretical distance $d$ is a function of $i$ and $k$. The higher order moments follow analogously to Eq.~(\ref{eq:partial_integration_math}) by definition [see Eq.~(\ref{eq:response_time}) and (\ref{eq:height})] as
\begin{eqnarray}
	\left<\tilde{t}\right> &=& \frac{d+1}{\lambda_1} \label{eq:tilde_mean}\\
	\tilde{\sigma} &=& \sqrt{\left<\tilde{t}^2\right> - \left<\tilde{t}\right>^2} = \sqrt{ \frac{d+1}{\lambda_1^2} } \label{eq:tilde_sigma}\\
	\tilde{H} &=& \frac{\tilde{Z}}{\tilde{\sigma}} = \frac{d!}{\lambda_1^d\sqrt{d+1}} \nonumber\\
	&\sim& \sqrt{2\pi} \left(\frac{d}{\lambda_1}\right)^d \exp(-d) \quad \mathrm{as} \quad d \rightarrow \infty\label{eq:tilde_H} \,,
\end{eqnarray}
where the last line denotes the asymptotic behavior for large distances $d \rightarrow \infty$ (see appendix B for detailed derivations).

\begin{figure}
\centering
\includegraphics[width=0.45\textwidth]{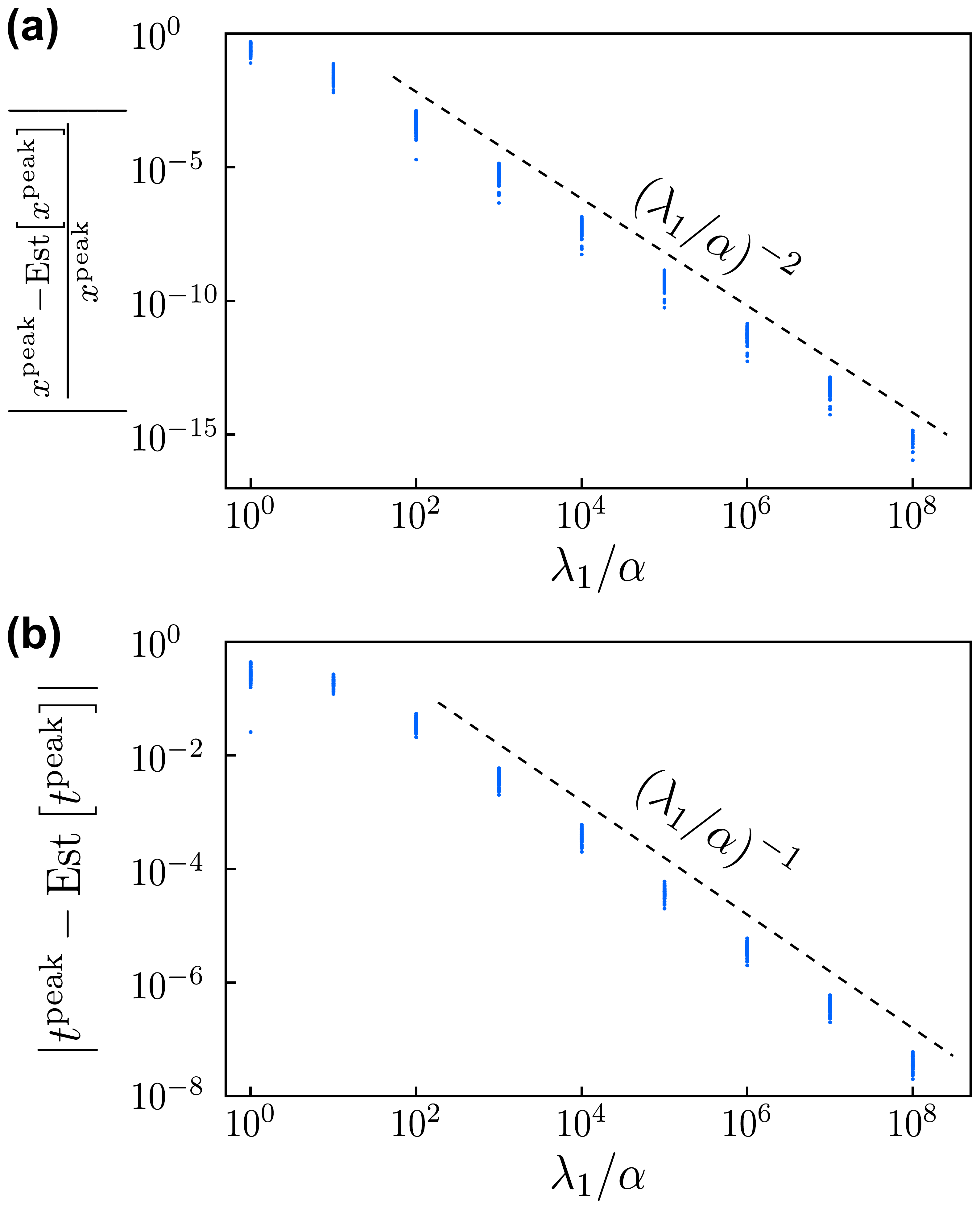}
\caption{\textbf{Exact estimators in the limit of weak coupling.} (a) Relative error of the estimated peak height Eq.~(\ref{eq:est_h}). (b) Absolute error of the estimated peak time Eq.~(\ref{eq:est_t}). 
Both errors disappear for large $\lambda_1/\alpha$, that means for weak coupling $\alpha/\lambda_1 \rightarrow 0$. Each point indicates one observation in $R=100$ connected Poisson random networks with $N=20$ units and $M=40$ links with fixed $\lambda_1 = \beta = 1$ and variable coupling strength $\alpha$, regardless of the distance of the unit to the origin of the perturbation.
}
\label{fig:weak_coupling_error}
\end{figure}

To obtain the peak response measures we determine the maximum $\frac{\mathrm{d}}{\mathrm{d}t} \tilde{x}(t)\mid_{t=\tpeaktilde} = 0$ and find
\begin{equation}
\tpeaktilde = \frac{d}{\lambda_1} \, \label{eq:approximation_tpeak}
\end{equation}
and consequently
\begin{equation}
\tilde{x}^\mathrm{peak} = \tilde{x}(\tpeaktilde) =  \left(\frac{d}{\lambda_1}\right)^d \exp(-d) \,. \label{eq:approximation_xpeak}
\end{equation}
Comparing Eq.~(\ref{eq:approximation_tpeak}) to (\ref{eq:tilde_mean}) and Eq.~(\ref{eq:approximation_xpeak}) to (\ref{eq:tilde_H}) suggests the bias corrections 
\begin{equation}
c_T = \tpeaktilde -\left<\tilde{t}\right> = - \frac{1}{\lambda_1} 
\end{equation}
and
\begin{eqnarray}
c_H &=& \frac{\tilde{x}^\mathrm{peak}}{\tilde{H}} = \frac{\sqrt{d+1} \, d^d}{\exp\left(d\right) \, d!} \nonumber\\
&=& \frac{1}{\sqrt{2\pi}} + \mathcal{O}(d^{-1}) \,,
\end{eqnarray}
where the last line describes the asymptotic behavior for large distances $d$. Note that these factors are independent of the origin of the perturbation $k$ or the specific unit $i$ for large distances but only depend on the network structure through the largest eigenvalue $-\lambda_1$. Analogously, we now convert the characteristic response measures for the original trajectories $x_i(t)$ to the estimators for the peak height and position
\begin{eqnarray}
\mathrm{Est}\left[\tpeak\right]_i &=& \left<t\right>_i - \frac{1}{\lambda_1} = -\frac{\left( M^{-2} \mathbf{x_0} \right)_{i}}{\left( M^{-1} \mathbf{x_0} \right)_{i}} - \frac{1}{\lambda_1} \label{eq:est_t}\\
\mathrm{Est}\left[x^\mathrm{peak}\right]_i &=& \frac{\sqrt{d+1} \, d^{d}}{\exp\left(d\right) \, d!} \, H_i \label{eq:est_h} \\
										 &=& \frac{H_i}{\sqrt{2\pi}} + \mathcal{O}(d^{-1})\,	\nonumber
\end{eqnarray}
where $d = d(i,k)$ is the graph theoretical distance from the perturbed unit $k$ to unit $i$. 
Consequently, this also suggests the new definition of the typical response duration as $\Delta t_i = Z_i / \mathrm{Est}\left[x(\tpeak)\right]_i = \sqrt{2\pi} \, \sigma_i + \mathcal{O}(d^{-1})$, illustrated in Fig.~\ref{fig:estimators}. 
\begin{figure*}
\centering
\includegraphics[width=0.9\textwidth]{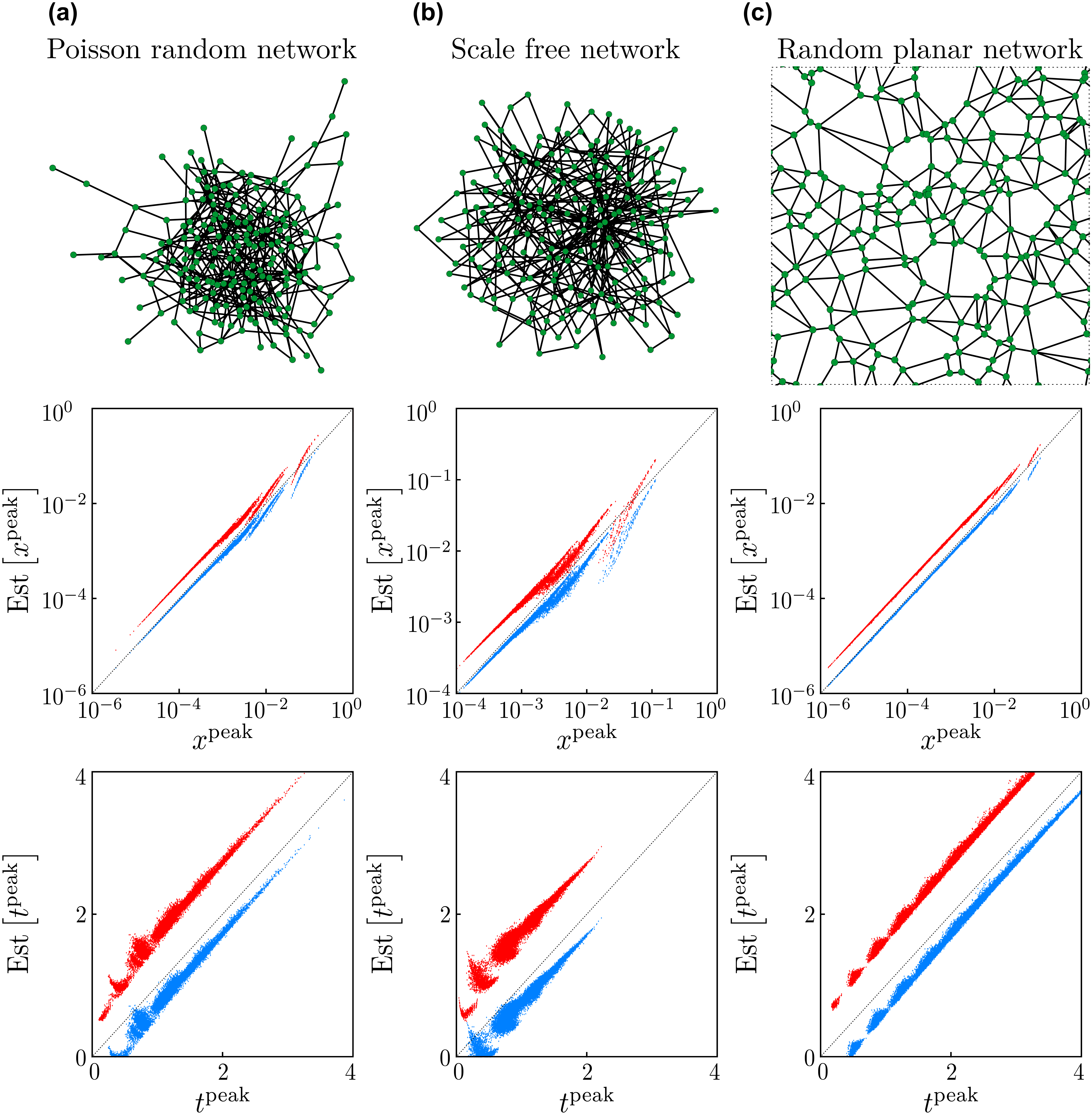}
\caption{\textbf{Improved estimation of peak values across network topologies.} 
The top row shows examples of the network topologies of (a) Poisson random networks, (b) Barabasi-Albert scale free random networks and (c) geometrically embedded random networks with periodic boundaries. The middle row shows the resulting $\mathrm{Est}\left[x(\tpeak)\right]$ [blue, Eq.(\ref{eq:est_h})] versus the true peak height $\xpeak$, the bottom row shows the corresponding results for $\mathrm{Est}\left[\tpeak\right]$ [blue, Eq.(\ref{eq:est_t})]. Points on the diagonal indicate perfect agreement of the estimated peak time or height with the actual peak time or height. 
Both estimators improve the prediction of the actual peak values compared to the raw expectation values (red). In both cases the estimators become more accurate for larger distances (smaller $\xpeak$ and larger $\tpeak$). 
All networks consist of $N=200$ units with $E = 400$ undirected interactions. The simulation parameters are $\beta = \lambda_1 = 1$ and $\alpha / \lambda_1 = 1$ in all three cases. The plots shows results for $R=10$ different realizations of the network structure where every unit was perturbed once, for a total of $400000$ measurements.}
\label{fig:networks}
\end{figure*}

\begin{figure*}
\centering
\includegraphics[width=0.9\textwidth]{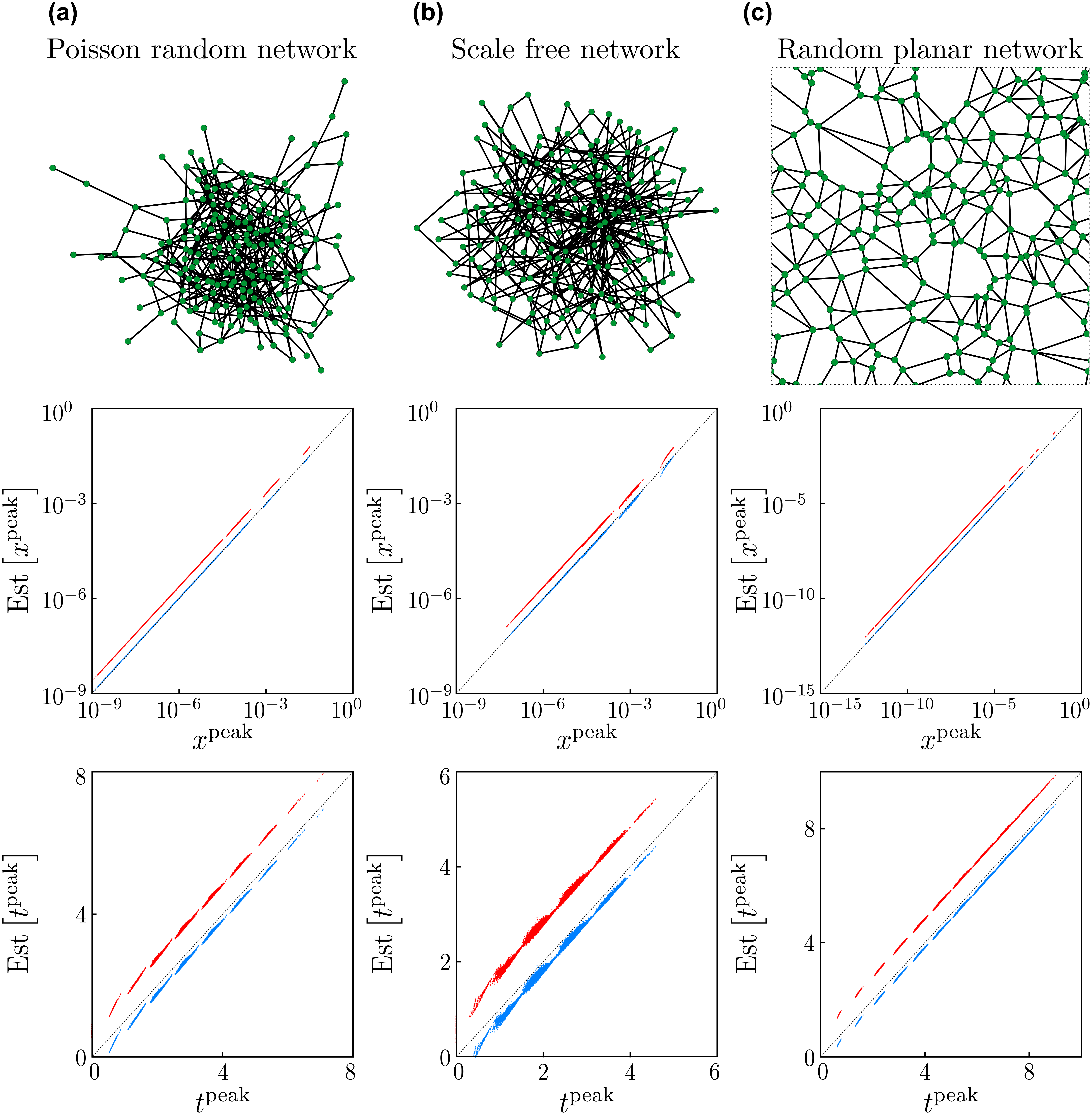}
\caption{\textbf{Accurate estimation of peak values across network topologies with weak coupling.} The top row shows examples of the network topologies of (a) Poisson random networks, (b) Barabasi-Albert scale free random networks and (c) geometrically embedded random networks with periodic boundaries. The middle row shows the resulting $\mathrm{Est}\left[x(\tpeak)\right]$ [blue, Eq.(\ref{eq:est_h})] versus the true peak height $\xpeak$, the bottom row shows the corresponding results for $\mathrm{Est}\left[\tpeak\right]$ [blue, Eq.(\ref{eq:est_t})]. Points on the diagonal indicate perfect agreement of the estimated peak time or height with the actual peak time or height. 
Both estimators more accurately predict the actual peak values compared to the raw expectation values (red). In both cases the estimators become more accurate for larger distances (smaller $\xpeak$ and larger $\tpeak$). 
All networks consist of $N=200$ units with $E = 400$ undirected interactions. The simulation parameters are $\beta = \lambda_1 = 1$ and $\alpha / \lambda_1 = 0.1$ in all three cases. The estimators are more accurate for this weaker coupling (compared to those in Fig.~\ref{fig:networks} for $\alpha = 1$). The plots show results for $R=10$ different realizations of the network structure where every unit was perturbed once, for a total of $400000$ measurements.}
\label{fig:networks_alpha_01}
\end{figure*}

\clearpage

\section{Universal ballistic spreading for weak coupling?}
 
The calculations above do not mathematically imply 
that these results should extend to the real response dynamics. Specifically, the limiting behavior of $\tilde{x}(t)$ is not purely exponential for $t \rightarrow \infty$ but scales as $t^d \exp\left(-\lambda_1 t\right)$. Beyond numerical validation of the results, we compute the same adjustments for other approximating functions that \emph{do} exhibit the correct asymptotic scaling for both $t\rightarrow 0$ and $t\rightarrow \infty$. Importantly, comparing the results of all of these calculations, we find \emph{identical} adjustments as above in the limit of weak coupling $\alpha/\lambda_1 \rightarrow 0$ (see appendix C for details and calculations). Numerical analysis, illustrated in Fig.~\ref{fig:weak_coupling_error}, supports that estimator errors indeed decay to zero for weak coupling $\alpha/\lambda_1 \rightarrow 0$. Together, these results suggest that the adjustments we derived above are universal 
in this limit. 

Moreover, both the peak response time as well as the characteristic response times increase linearly with the distance $d$ in this limit for all families of approximating response functions. This indicates that the spreading of the perturbation is ballistic, even though the coupling is diffusive. This observation is in line with heuristic predictions for different dynamics such as diseases spreading in transportation networks \cite{gautreau07_arrivaltimesDisease, gautreau08_diseaseSpread, brockmann13_spreading, Iannelli17_effectiveDistances, chen18_arrivalTimesLinearSpreading}. In these models the mobility rate (coupling strength) is typically much slower then the internal dynamics of the individual units (weak coupling limit) and the observed arrival time increases linearly with the (effective) distance of a unit to the original outbreak location.

\section{Accurate estimation across network topologies}

We numerically test the accuracy of the estimators across different network topologies for fixed $\alpha/\lambda_1$. We perturb each unit in the network once and record the resulting typical response times and magnitudes as well as peak values. For simplicity we use constant coupling strengths $\alpha_{ij} \in \left\{0, \alpha\right\}$ and identical internal dynamics $\beta_i = \beta = \lambda_1 = 1$ in these examples. However, this is not a necessary condition for our results to hold as the derivation given above holds for general matrices $M$, assuming only $x_i(t) > 0$ for all $t > 0$ and $x_i(t) \rightarrow 0$ as $t \rightarrow \infty$.

Figure~\ref{fig:networks} shows the results for Poisson random networks (narrow degree distribution, small diameter), Barabasi-Albert scale free random networks (broad degree distribution, small diameter) and random geometrically embedded networks (narrow degree distribution, large diameter) for $\alpha/\lambda_1 = 1$. The adjustment systematically improves the estimate compared to the characteristic response values but is still not exact, as expected for non-zero $\alpha/\lambda_1$. Specifically, the peak time is typically underestimated. The estimate of the peak height becomes more accurate for large distances (small $\xpeak$). Figure~\ref{fig:networks_alpha_01} shows the same simulations with weaker coupling $\alpha/\lambda_1 = 0.1$. As expected from the analytical calculations, the estimates agree much better with the exact peak values. Additional results for absolute and relative errors of the estimators are shown in appendix A.

We specifically note, that all assumptions in the derivations presented above are satisfied also for directed networks or networks with heterogeneous coupling strengths. As seen in Fig.~\ref{fig:networks_alpha_01}(b), heterogeneous network structures (and similarly heterogeneous coupling strengths) cause larger fluctuations in the estimations. The reason is the existence of multiple short paths or stronger coupling along these paths in such networks. However, the analytic results remain correct. Most importantly, the estimators become exact in the limit of weak coupling, independent of the network topology or coupling strength distribution.

\section{Conclusion}

Understanding the propagation of perturbation-induced signals in networked systems helps to predict, control and mitigate their impact in a range of processes in biology and engineering, from epidemic spreading of diseases to the impact of load shedding or infrastructure outages in electric power grids. Among the fundamental questions are when and how strongly perturbations initiated at some unit in a network reach other units. So far, it has been impossible to analytically estimate timing and strengths of such signals as an explicit function of the underlying base state of the system and the network's interaction topology. These limitations hold even for linear deterministic systems because the equations determining peak timing and strength are transcendental and as such mathematically intractable.

A recent proposal \cite{wolter18_expectationSpreading} suggests to take a complementary perspective and predicts characteristic arrival times and strengths not in terms of peak times and amplitudes but in terms of expectation values that result from interpreting the deterministic trajectory of a unit's  response as a probability density. This approach yields characteristic arrival times and strengths as explicit functions of the inverse of the Jacobian matrix that in turn encodes both the base operating state and the interaction topology. However, these characteristic quantities are not intended to predict peak times and amplitudes -- and if interpreted as such, exhibit large errors. So it still remains unclear how to explicitly quantify peak times and amplitudes.

Here we connect the two sets of quantifiers and derive approximate analytical estimators for the absolute peak positions and heights of the responses in terms of quantifiers based on expectation values. We employ qualitative approximations of the response functions mimicking the asymptotic behavior both for small and large times. The resulting estimators enable approximate predictions of the peak timings and heights across network topologies. Interestingly, in the weak coupling regime (see appendix C), i.e. asymptotically as $\alpha / \lambda_1\rightarrow 0$, the predictions become identical across all specific approximating functions tested, suggesting universality. Outside the asymptotic regime, i.e. for stronger coupling, the adjusted estimators seem to systematically underestimate the peak response values.

Together with the expressions for the characteristic response measures derived by Wolter et al. in terms of expectation values \cite{wolter18_expectationSpreading}, these results provide an analytic framework for predicting the impact of perturbations on any unit in any network operating close to a stable fixed point in the limit of weak coupling. Our results on deterministic systems are thereby complementing the analyses for specific models of disease spreading \cite{gautreau07_arrivaltimesDisease, gautreau08_diseaseSpread, brockmann13_spreading, Iannelli17_effectiveDistances, chen18_arrivalTimesLinearSpreading}. They moreover suggest that in the asymptotic regime of weak coupling, perturbations spread ballistically through the network, even though the coupling is diffusive. Further work must show how details of the local network topology affect the accuracy of the predictions and how the results can be extended to allow also accurate predictions for stronger coupling and at close distances.\\

\section*{Acknowledgments}
We thank Raoul Schmidt and Benajmin Friedrich for helpful discussions.
This work was partially supported by the German Science Foundation (DFG) through the Cluster of Excellence `Center for Advancing Electronics Dresden' (cfaed) and the Federal Ministry of Education and Research (BMBF Grant No. 03SF0472F und 03EK3055F).

\bibliography{manuscript_PerturbationEstimator}

\onecolumngrid

\newpage

\section*{Appendix A:\\
Errors of the estimation}

\begin{figure*}[h!]
\centering
\includegraphics[width=0.8\textwidth]{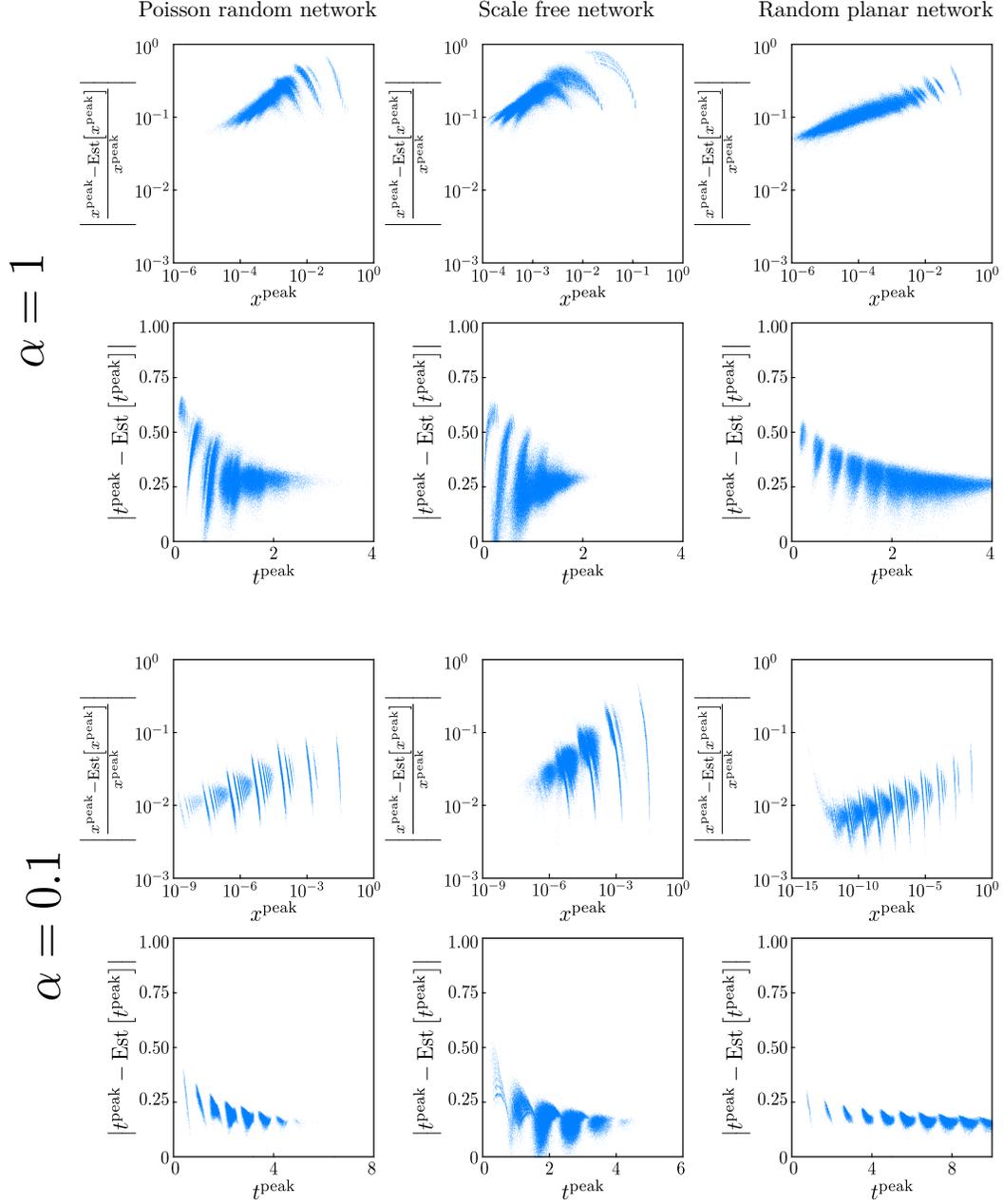}
\caption{\textbf{Errors of the estimation of peak height and time.} The figures show the relative error of the estimated peak height [main manuscript Eq.~(19)] (exponential scaling with distance) and the absolute error of the estimated peak time [main manuscript Eq.~(18)] (linear scaling with distance). The top part shows the error for coupling strength $\alpha/\lambda_1 = 1$ (compare Fig.~5 in the main manuscript), the bottom part shows the error for weaker coupling $\alpha/\lambda_1 = 0.1$ (compare Fig.~6 in the main manuscript) for the three different network toplogies, Poisson random networks (left), Barabasi-Albert scale free random networks (middle) and geometrically embedded random networks with periodic boundarie (right). All networks consist of $N=200$ units with $E = 400$ undirected interactions. The internal dynamics of the units is described by $\beta = \lambda_1 = 1$ in all cases. The estimators become more accurate for weaker coupling and larger distances. Each plot shows results for $R=10$ different realizations of the network structure where every unit was perturbed once, for a total of $400000$ measurements.}
\label{fig:networks_error}
\end{figure*}

\clearpage

\section*{Appendix B:\\
Detailed calculations for $\tilde{x}(t)$}

Here we provide details on the calculation of the characteristic and peak response measures for $\tilde{x}(t) = t^d \exp\left(-\lambda_1 t\right)$ in the main text. For the characteristic response values, we first calculate the total response
\begin{eqnarray}
\tilde{Z} &=& \int_0^\infty t^d \exp(  -\lambda_1 t) \mathrm{d}t \nonumber\\
	&=& \underbrace{\left[ d t^{d-1} \exp(  -\lambda_1 t) \right]_0^\infty}_{=0 \quad\mathrm{if}\quad d \ge 1} \nonumber\\
	&\null& \; + \int_0^\infty \frac{d}{ \lambda_1} t^{d-1} \exp( -\lambda_1 t) \mathrm{d}t \nonumber\\
	&\vdots& \nonumber \\
	&=& \left[ \frac{d!}{\lambda_1^d} \exp( -\lambda_1 t) \right]_0^\infty \nonumber\\
	&=& \frac{d!}{\lambda_1^{d+1}}	\,.
\end{eqnarray}
by repeated partial integration. We use this to define $\tilde{\rho}(t) = \tilde{x}(t) / \tilde{Z}$ and calculate the expectation values analogously:
\begin{eqnarray}
	\left<\tilde{t}\right> &=& \frac{1}{\tilde{Z}} \int_0^\infty t \tilde{x}(t) \mathrm{d}t \nonumber \\
						   &=& \frac{1}{\tilde{Z}} \int_0^\infty t^{d+1} \exp(  -\lambda_1 t) \mathrm{d}t \nonumber\\
						   &=& \frac{1}{\tilde{Z}} \underbrace{\left[ \left(d+1\right) t^{d} \exp(  -\lambda_1 t) \right]_0^\infty}_{=0} \nonumber\\
	&\null& \; + \frac{1}{\tilde{Z}} \int_0^\infty \frac{d+1}{ \lambda_1} t^{d} \exp( -\lambda_1 t) \mathrm{d}t \nonumber\\
	&\vdots& \nonumber \\
	&=& \frac{1}{\tilde{Z}}  \left[ \frac{\left(d+1\right)!}{\lambda_1^{d+1}} \exp( -\lambda_1 t) \right]_0^\infty \nonumber\\
	&=& \frac{1}{\tilde{Z}} \frac{(d+1)!}{\lambda_1^{d+2}} = \frac{d+1}{\lambda_1} \,, \label{eq:tilde_mean_app}
\end{eqnarray}
\begin{eqnarray}
   \left<\tilde{t}^2\right> &=& \frac{1}{\tilde{Z}} \int_0^\infty t^2 \tilde{x}(t) \mathrm{d}t \nonumber \\
						   &=& \frac{1}{\tilde{Z}} \underbrace{\left[ \left(d+2\right) t^{d+1} \exp(  -\lambda_1 t) \right]_0^\infty}_{=0} \nonumber\\
	&\null& \; + \frac{1}{\tilde{Z}} \int_0^\infty \frac{d+2}{ \lambda_1} t^{d+1} \exp( -\lambda_1 t) \mathrm{d}t \nonumber\\
	&\vdots& \nonumber \\
	&=& \frac{1}{\tilde{Z}}  \left[ \frac{\left(d+2\right)!}{\lambda_1^{d+2}} \exp( -\lambda_1 t) \right]_0^\infty \nonumber\\
	&=& \frac{1}{\tilde{Z}} \frac{(d+2)!}{\lambda_1^{d+3}} = \frac{\left(d+2\right)\left(d+1\right)}{\lambda_1^2} \,,	
\end{eqnarray}
and finally the standard deviation $\tilde{\sigma}$ as
\begin{eqnarray}
\tilde{\sigma} &=& \sqrt{\left<\tilde{t}^2\right> - \left<\tilde{t}\right>^2} \nonumber\\
	&=& \sqrt{\frac{(d+2)(d+1)}{\lambda_1^2} - \left(\frac{d+1}{\lambda_1}\right)^2} = \sqrt{\frac{ d+1 }{\lambda_1^2}} \,.
\end{eqnarray}
This results in the characteristic response magnitude 
\begin{equation}
\tilde{H} = \frac{\tilde{Z}}{\tilde{\sigma}} = \frac{d!}{\lambda_1^d\sqrt{d+1}} \label{eq:tilde_H_app}
\end{equation}
which with the asymptotic Stirling-approximation becomes
\begin{eqnarray}
\tilde{H} &\sim& \sqrt{2\pi} \left(\frac{d}{\lambda_1}\right)^d \exp(-d) \quad \mathrm{as} \quad d \rightarrow \infty \,.
\end{eqnarray}

Similarly, we calculate the peak response values by finding the maximum of $\tilde{x}(t)$ by solving $\mathrm{d} \tilde{x}(t) / \mathrm{d} t \mid_{t = \tpeaktilde} = 0$:
\begin{eqnarray}
\frac{\mathrm{d}}{\mathrm{d}t} \tilde{x}(t)\mid_{t = \tpeaktilde} &=& d \left(\tpeaktilde\right)^{d-1} \exp\left(-\lambda_1 \,\tpeaktilde\right) \nonumber\\
 &\null& - \lambda_1 \left(\tpeaktilde\right)^d \exp\left(-\lambda_1 \,\tpeaktilde\right) = 0 \nonumber\\
\Rightarrow  \quad  \tpeaktilde &=& \frac{d}{\lambda_1} \, \label{eq:approximation_tpeak_app}
\end{eqnarray}
which directly leads to the peak response amplitude
\begin{equation}
\tilde{x}^\mathrm{peak} = \tilde{x}(\tpeaktilde) = \frac{d^d \exp(-d)}{\lambda_1} \,. \label{eq:approximation_xpeak_app}
\end{equation}

Comparing these results [Eq.~(\ref{eq:tilde_mean_app}) with Eq.~(\ref{eq:approximation_tpeak_app}) and Eq.~(\ref{eq:tilde_H_app}) with Eq.~(\ref{eq:approximation_xpeak_app})] leads to the constants $c_T$ and $c_H$ as
\begin{equation}
c_T = \tpeaktilde -\left<\tilde{t}\right> = - \frac{1}{\lambda_1} 
\end{equation}
and
\begin{eqnarray}
c_H &=& \frac{\tilde{x}^\mathrm{peak}}{\tilde{H}} = \frac{\sqrt{d+1} \, d^d}{\exp\left(d\right) d!} \nonumber\\
&=& \frac{1}{\sqrt{2\pi}} + \mathcal{O}(d^{-1}) \,,
\end{eqnarray}
and the estimators given in the main text in Eq.~(18) and (19) in the main manuscript.

\newpage

\section*{Appendix C:\\
Other approximating functions}

Here we consider other functional forms of $\tilde{x}(t)$ as approximation models. We follow the same arguments as above, first determining the correct parameters and then calculating the characteristic and peak response values. We assume that the response factors into two parts, $\tilde{x}(t) = f(t) g(t)$, with the following conditions:  we capture the behavior at small $t$ in the function $f(t)$ with $f(t) \sim t^d$ and $g(0) = 1$ for $t \rightarrow 0$. For large times $t \rightarrow \infty$, we have $f(t) \rightarrow c_1$ and $g(t) \sim \exp\left( -\lambda_1 t \right)$, where $c_1$ is the constant coefficient of the expansion in terms of eigenvalues [see Eq.~(9)].

\subsection*{Case (i)}

As the second approximating function we consider
\begin{equation}
\tilde{x}(t) = c_1 \left[1 - \exp\left({-\gamma t}\right)\right]^d \exp\left(-\lambda_1 t\right) \,,
\end{equation}
such that $f(t) = c_1 \left[1 - \exp\left({-\gamma t}\right)\right]^d$ and $g(t) = \exp\left(-\lambda_1 t\right)$. This leads to the correct asymptotic behavior as
\begin{eqnarray}
	x_i(t) &=& c_1 \gamma^d t^d + \mathcal{O}(t^{d+1}) \quad\mathrm{as}\quad t\rightarrow 0\\
	x_i(t) &\sim& c_1 \exp\left(-\lambda_1 t\right) \quad\mathrm{as}\quad t\rightarrow \infty\,.
\end{eqnarray}
For undirected networks with homogeneous parameters $\alpha$ and $\beta$ the eigenvector $\mathbf{v}_1$ to the largest eigenvalue $-\lambda_1 = - \beta$ is $\mathbf{v}_1 = (1,1, \dots, 1)^T$ and we have $c_1 = 1/N$. We then find that $\gamma = \alpha N^{1/d}$ directly proportional to the coupling strength $\alpha$ leads to the correct asymptotic behavior $\alpha^d t^d$ for $t\rightarrow 0$. 

Direct calculation then yields the characteristic and peak response values
\begin{eqnarray}
\left<\tilde{t}\right> &=& \frac{\mathrm{PG}[0,1+\lambda_1/\gamma+d] - \mathrm{PG}[0,1+\lambda_1/\gamma]}{\gamma} \nonumber\\
			&\sim& \frac{d+1}{\lambda_1} \quad\mathrm{as}\quad \frac{ \gamma }{\lambda_1} \rightarrow 0 \\
\tilde{t}^\mathrm{peak} &=& \frac{\log\left(1 + \gamma d / \lambda_1\right)}{\gamma} \nonumber\\
			&\sim& \frac{d}{\lambda_1} \quad\mathrm{as}\quad \frac{ \gamma }{\lambda_1} \rightarrow 0
\end{eqnarray}
for the response times and
\begin{eqnarray*}
\tilde{H} &=& \frac{d! \Gamma[\lambda_1/\gamma]}{\Gamma[1+\lambda_1/\gamma+d]} \\
	& & \quad \times \frac{1}{\sqrt{ \mathrm{PG}[1,1+\lambda_1/\gamma] - \mathrm{PG}[1,1+\lambda_1/\gamma +d] }} \nonumber\\
	&\sim& \frac{\gamma^d \, d!}{\lambda_1^d \, \sqrt{d+1}} \quad\mathrm{as}\quad \frac{ \gamma }{\lambda_1} \rightarrow 0 \\
\tilde{x}^\mathrm{peak} &=& \left(\frac{\gamma d}{\lambda_1 + \gamma d}\right)^d \left(1 + \frac{\gamma d}{\lambda_1}\right)^{-\lambda_1 / \gamma} \nonumber\\
	&\sim& \frac{\gamma^d \, d^d}{\lambda_1^d} \exp\left(-d\right) \quad\mathrm{as}\quad \frac{ \gamma }{\lambda_1} \rightarrow 0
\end{eqnarray*}
for the response magnitude, where $\mathrm{PG}[n,z]$ denotes the PolyGamma function, the $n$-th derivative of the digamma function $\mathrm{PG}[0,z] = \frac{1}{\Gamma[z]} \frac{\mathrm{d}\Gamma[z]}{\mathrm{d}z}$. The second line for each equation denotes the leading order of the asymptotic behavior for weak coupling $\alpha/\lambda_1 \sim \gamma/\lambda_1 \rightarrow 0$. We then obtain the same relation between the characteristic and peak response values as with the approximation in the main text
\begin{equation}
 c_T = \tpeaktilde - \left<\tilde{t}\right> \sim \frac{1}{\lambda_1}
\end{equation}
and
\begin{eqnarray}
 c_H &=& \frac{\tilde{x}^\mathrm{peak}}{\tilde{H}} \sim  \frac{\sqrt{d+1} \, d^d}{\exp\left(d\right) \, d!} \nonumber\\
 &\sim& \frac{1}{\sqrt{2\pi}} \,
\end{eqnarray}
where the first line holds in the limit of weak coupling $\alpha/\lambda_1 \sim \gamma/\lambda_1 \rightarrow 0$ and the second line holds if additionally $d \rightarrow \infty$.

\subsection*{Case (ii)}

Third, we consider the approximation
\begin{equation}
\tilde{x}(t) = c_1 \left(\frac{\gamma t}{1 + \gamma t}\right)^d \exp\left(-\lambda_1 t\right) \,,
\end{equation}
with $f(t) = c_1 \left(\frac{\gamma t}{1 + \gamma t}\right)^d$ and $g(t) = \exp\left(-\lambda_1 t\right)$. This leads to the correct asymptotic behavior as in case (i),
\begin{eqnarray}
	x_i(t) &=& c_1 \gamma^d t^d + \mathcal{O}(t^{d+1}) \quad\mathrm{as}\quad t\rightarrow 0\\
	x_i(t) &\sim& c_1 \exp\left(-\lambda_1 t\right) \quad\mathrm{as}\quad t\rightarrow \infty\,.
\end{eqnarray}
For undirected networks with homogeneous parameters $\alpha$ and $\beta$ the eigenvector $\mathbf{v}_1$ to the largest eigenvalue $-\lambda_1 = - \beta$ is $\mathbf{v}_1 = (1,1, \dots, 1)^T$ and we have $c_1 = 1/N$. We then find that $\gamma = \alpha N^{1/d}$ directly proportional to the coupling strength $\alpha$ leads to the correct asymptotic behavior $\alpha^d t^d$ for $t\rightarrow 0$.

Writing $\delta = \lambda_1 / \gamma$ for brevity, direct calculation yields the characteristic and peak response values
\begin{eqnarray}
\left<\tilde{t}\right> &=& \frac{d+1}{\lambda_1} \times \frac{\mathrm{HGU}[d,-1,\delta]}{\mathrm{HGU}[d,0,\delta]} \nonumber\\
			&\sim& \frac{d+1}{\lambda_1} \quad\mathrm{as}\quad \frac{ \gamma }{\lambda_1} \rightarrow 0\\
\tilde{t}^\mathrm{peak} &=& \frac{-\delta + \sqrt{\delta^2 + 4 d \delta}}{2\lambda_1} \nonumber\\
			&\sim& \frac{d}{\lambda_1} \quad\mathrm{as}\quad \frac{ \gamma }{\lambda_1} \rightarrow 0
\end{eqnarray}
for the response times and
\begin{eqnarray}
H &=& \frac{d! \, \mathrm{HGU}[d,-1,\delta]^{-1}}{ \sqrt{ \frac{d+2}{d+1} \frac{\mathrm{HGU}[d,-2,\delta] \mathrm{HGU}[d,0,\delta]}{\mathrm{HGU}[d,-1,\delta]^2 } - 1} } \nonumber\\
	&\sim& \frac{\gamma^d \, d!}{\lambda_1^d \, \sqrt{d+1}} \quad\mathrm{as}\quad \frac{ \gamma }{\lambda_1} \rightarrow 0 \\
x^\mathrm{peak} &=& \left[1 - \exp\left(1/2 - 1/2 \sqrt{1 + 4 d \gamma / \lambda_1}\right)\right]^d \\
			& & \quad \times \exp\left(\delta - \sqrt{\delta^2 + 4 d \delta}\right) \nonumber \\
&\sim& \frac{\gamma^d \, d^d}{\lambda_1^d} \exp\left(-d\right) \quad\mathrm{as}\quad \frac{ \gamma }{\lambda_1} \rightarrow 0
\end{eqnarray}
for the response magnitude, where $HGU$ denotes Tricomi's confluent hypergeometric function
\begin{equation*}
\mathrm{HGU}[a,b,z] = \frac{1}{\Gamma[a]} \int_0^\infty e^{-z t} t^{a-1} \left(1+t\right)^{b-a-1} \mathrm{d}t \,.
\end{equation*}
The second line for each equation denotes the leading order of the asymptotic behavior for weak coupling $\alpha/\lambda_1 \sim \gamma/\lambda_1 \rightarrow 0$. We again obtain the same relation between the characteristic and peak response values as with the approximation in the main text
\begin{equation}
 c_T = \tpeaktilde - \left<\tilde{t}\right> \sim \frac{1}{\lambda_1}
\end{equation}
and
\begin{eqnarray}
 c_H &=& \frac{\tilde{x}^\mathrm{peak}}{\tilde{H}} \sim  \frac{\sqrt{d+1} \, d^d}{\exp\left(d\right) \, d!} \nonumber\\
 &\sim& \frac{1}{\sqrt{2\pi}} \,
\end{eqnarray}
where the first line holds in the limit of weak coupling $\alpha/\lambda_1 \sim \gamma/\lambda_1 \rightarrow 0$ and the second line holds if additionally $d \rightarrow \infty$.

\end{document}